\documentclass[11pt]{article}

\usepackage[T1]{fontenc}
\usepackage{lmodern}
\usepackage[margin=1in]{geometry}
\usepackage{setspace}
\usepackage{amsmath,amssymb}
\usepackage{amsthm}
\usepackage{graphicx}
\usepackage{xcolor}
\usepackage{booktabs}
\usepackage{float}
\usepackage{caption}
\usepackage{url}
\usepackage{hyperref}
\hypersetup{
  colorlinks=true,
  linkcolor=black,
  citecolor=black,
  urlcolor=black
}
\usepackage[authoryear,round]{natbib}

\theoremstyle{plain}
\newtheorem{proposition}{Proposition}

\title{The Price of Submission\thanks{I thank Noah Macdonald for useful comments on an earlier version of the paper. This paper was created with the help of Anthropic's Claude Code (Opus 4.8), OpenAI's Codex (GPT-5.5) and refine.ink; these tools assisted with data collection, derivations, drafting and computational checks. I reviewed every claim and remain responsible for all errors. Cite this paper as: Fourie, Johan. 2026. ``The Price of Submission.'' Working Paper, Department of Economics, Stellenbosch University.}}
\author{Johan Fourie\thanks{Department of Economics, Stellenbosch University.
Email: \href{mailto:johanf@sun.ac.za}{johanf@sun.ac.za}.}}
\date{}

\begin{document}

\maketitle

\begin{abstract}
\noindent Economics journals ration access to peer review with time and money. A July 2026 census of 500 RePEc-ranked economics journals accepting unsolicited submissions finds that 113 (22.6 per cent) charge a fee, with incidence falling from 68 per cent in the top 50 to 2 per cent in the bottom 50. A sequential-submission model shows that waiting pays part of the capacity-clearing price, leaving the fee as a residual. Because rejected manuscripts move down the ranking, a fee reduces system-wide referee demand only when authors stop, shorten their submission chains or switch to journals requiring fewer reports.
\end{abstract}

\vspace{1em}\noindent\textbf{Keywords:} scientific publishing; submission fees; peer review; congestion; artificial intelligence

\vspace{0.25em}

\vspace{0.5em}

\newpage

\section{Introduction}\label{sec:intro}

Economics journals ration access to certification with time. An author may submit a manuscript to one journal at a time. A rejection therefore delays the next attempt by months. Review itself delays any eventual return from publication. A submission fee adds a money price and can finance more refereeing. The two instruments do different things. Delay uses up the author's time and slows the movement of rejected manuscripts down the ranking. A fee deters entry without itself delaying the next attempt. When new manuscripts multiply faster than the supply of careful review, the question is not what one journal should charge. It is how the whole ranked system prices one more attempt and where the attempts it discourages end up.

This paper combines a measurement exercise with a model of that system. The measurement is a census: submission fees, waivers, article-processing charges and referee-payment policies for the 500 highest-ranked economics journals in the July 2026 RePEc aggregate ranking that accept unsolicited submissions. The collection is automated. Automation buys completeness and a reproducible trail of evidence. I keep three things about each record separate: whether the journal charges for submission, whether the amount is established and whether the evidence is a retrieved statement or only silence. The model then follows a manuscript through a sequence of journals. At each journal the author weighs the expected gain from trying it against sending the paper straight to the next one. Waiting shrinks that gain. Submissions that survive the editor's initial screen, called desk triage, go out to referees. Capacity is therefore counted in completed referee reports rather than in submissions.

Three findings organise the paper. First, fees are common at the top of the profession and rare below it. Of the 500 journals in the frame, 113 charge for submission, which is 22.6 per cent. Treating all 7 unresolved records as chargers raises the classification bound to 24.0 per cent. In the highest-ranked fiftieth of the frame 68 per cent charge a fee. In the lowest-ranked fiftieth 2 per cent do. Second, nominal fees mostly rose among the positive fees that are comparable in the matched historical sample. Among 66 journals matched to a 2019 compilation, 23 raised a positive fee in the same posted currency, one lowered it, two removed it and one introduced an ordinary fee. Five required membership in 2019 and cannot be classified as zero-dollar submission. Third, the model separates the full price of entry from its money part. A journal that rations entry with a long wait needs a smaller fee to ration the same queue. Whether a fee helps the profession rather than one journal depends on what the turned-away authors do next.

The report-unit accounting also clarifies what fee revenue can finance. A submission fee collects revenue from each non-waived submission. Only some submissions pass desk triage. Each manuscript that goes out for review needs several completed reports. Fee revenue per submission is therefore not the same thing as a possible payment per report. Under the counterfactual in which a journal hands all fee revenue to its referees, the payment per report divides total revenue by total completed reports. It is below the fee only when the average submission generates more than one completed report. Whether payment would actually expand review capacity depends on how referee effort responds to it. A cross-section of posted fees cannot reveal that response.

Recent work documents a publishing system under strain: article output has grown faster than the population available to vet it, while turnaround times vary widely across publishers \citep{hanson2024strain}. Delays are particularly long in economics and business, and exclusive submission means that rejection can postpone every later attempt \citep{bjork2013delay}. Peer review also uses heterogeneous effort. \citet{zhang2022endeavor} use report length as a proxy for that effort, while \citet{wolfram2021exploration} find in their sample of open-review journals that additional reviewers, reports and revision rounds are generally not associated with more citations. These findings motivate counting capacity in completed reports rather than submissions, but they do not explain how a ranked system rations attempts. \citet{zollman2024journal} model how a journal's incentives shape the acceptance standard it sets in peer review, a within-journal margin that complements the cross-journal routing studied here. The model builds on work that treats submission fees and response times as joint design margins \citep{cotton2013submission}, studies authors' sequential submission strategies \citep{heintzelman2009where} and explains the supply of referee effort \citep{engers1998referees,chetty2014prosocial}. It combines the continuation value from sequential-submission models with the physical conversion of submissions into referee reports. A fee at a top journal can move a manuscript rather than remove it.

A parallel literature studies author-facing prices. \citet{schonfelder2020charges} relates article-processing charges to journal impact, publisher and field. \citet{klebel2023barrier} associates those charges with stratification by institutional resources, while \citet{oldford2023marginalizing} finds that submission fees narrow the opportunity set of less-resourced accounting and finance authors. \citet{asai2024determinants} estimates relatively inelastic manuscript submissions to article-processing charges among fully open-access journals at two publishers. Article-processing charges are paid for publication; submission fees price an attempt. This distinction is central to the paper. Author prices also coexist with reader-side revenue and institutional subsidy \citep{jeon2010pricing,mccabe2018open}, while certification models explain why fee structure and rejection carry information \citep{farhi2013fear,pollrich2024irrelevance}. The census supplies the price schedule for entering review across a ranked discipline. It does not identify authors' price elasticity, the causal effect of fees on access or what journals maximise \citep{card2020editors}.

The rest of the paper is structured as follows. Section~\ref{sec:data} presents the census, its evidence audit and the historical comparison. Section~\ref{sec:model} develops the money-delay-continuation model and the routing result. Section~\ref{sec:quantitative} states what the available data identify and reports scenario maps. Section~\ref{sec:design} considers waivers, payment timing and AI-assisted triage. Section~\ref{sec:conclusion} concludes.

\section{Explicit prices in economics journals}\label{sec:data}

\subsection{Frame and collection}

The frame is the 500 highest-ranked economics journals to which an outside author can send an unsolicited manuscript. It is built from the RePEc/IDEAS aggregate journal ranking for July 2026, which combines seven citation-based metrics and is public and reproducible. The ranking indexes outlets, not only journals. Among its entries are conference proceedings volumes, commissioned review series, central-bank and international-organisation bulletins, an article repository and publications that have ceased. An author cannot submit a paper to any of them. I therefore remove those entries first and take the top 500 of what remains. Of the 600 entries collected, 55 are removed on this ground and 545 remain. The frame therefore reaches to raw rank 550. Building the frame in this order fixes its size at 500 journals. Had I instead cut the ranking at a fixed depth, the number of journals inside it would have depended on how many institutional series happened to fall there.

Each removal rests on the outlet's own pages. A mechanical rule nominates candidates from the RePEc archive code, from title patterns and from the absence of a registered identity or of recent publication. A search agent then reads the candidate's pages and reports whether it accepts unsolicited submissions, with the sentence it relied on. Nomination alone never removes an entry, because the rule errs in both directions: it nominated the \emph{Proceedings of the National Academy of Sciences} and all three series of the \emph{Journal of the Royal Statistical Society}, all of which take unsolicited work and all of which stay. The removals comprise 19 publications that have ceased, 18 institutional bulletins, 12 proceedings volumes, 5 commissioned review series and 1 repository. Appendix~\ref{sec:appendix} lists the rule and the counts.

Three considerations put the cut at 500 rather than deeper. First, the evidence thins with depth. The share of records whose zero rests on silence rather than on a retrieved statement rises across hundred-rank bands from 34 to 77 per cent. Second, identification as an economics outlet thins with it. The share of entries appearing in Scimago's economics, econometrics and finance list falls across the same bands from 77 to 41 per cent. Third, the ranking leaves the discipline. The 500th journal in the frame is still one an economist would target. Raw rank 550 is \emph{Socio-Economic Planning Sciences} and raw rank 600 is \emph{Policy Sciences}. The cut is not load-bearing. Every statistic below is also computed for all 545 eligible journals. On that longer list fee incidence is 21.1 per cent rather than 22.6 per cent. The gradient runs from 68 per cent in the first fifty journals to 4 per cent in the last fifty. The median fee is \$139 rather than \$143. Nothing in the paper turns on those differences.

For each entry a web-search agent consulted four sources in descending order of authority: author guidelines, the submission portal, publisher fee pages and society announcements. It returned the submission fee and its currency, member and student prices, waivers, refundability, page charges, article-processing charges, referee payment and the supporting evidence. Every record retains the source addresses, the retrieved text, a confidence grade and the collection date. The procedure produced a record for all 600 entries.

A second automated collection covered the top 100 entries and a 10 per cent random sample of the remainder, yielding 146 comparable records. The two collections disagree on whether a fee exists in five of them. That is 3.4 per cent of the sample, with a 95 per cent Wilson interval of 1.5 to 7.8 per cent. The sample is stratified by design and over-weights the top of the ranking. Weighting the two strata by their size in the collected ranking puts the disagreement rate at 3.9 per cent.

A blind hand check then covered all nine records contested on status, amount or page access across the automated collections and 20 journals drawn across rank bands, half coded as charging and half as not charging. It corrected two fee-status classifications and three amounts before the statistics below were computed.

\subsection{Evidence classes and current incidence}

Three things about a record are separate. The classification keeps them separate: whether the journal charges, what the price is and how good the evidence is. A journal charges when it levies a price on every submitted manuscript. A price is established when the retained text contains the amount. Evidence is strong when the text states the charge and weak when it rests on silence.

Two arrangements have to be told apart, because conflating them prices the wrong thing. Some journals charge a fee for each submission and also require society membership. \emph{Econometrica} is one: members pay \$125 for every submission. That is a per-submission price and it counts as one. Other journals charge nothing per submission but admit only members. The \emph{Journal of Risk \& Insurance} is one. ARIA membership is required in order to submit. Members then submit free. The annual dues also buy journal subscriptions. Those dues are not the price of one attempt. Counting them as such would price a year's subscription as the cost of a single submission. The test that separates the two is what a member pays to submit.

On these rules 113 journals in the frame charge for submission. For 112 of them the amount is established. For one the search did not recover it. One further journal admits only members. Of the rest, 42 carry retrieved text stating that no submission charge applies, 337 rest on the prescribed search finding none and 7 are unresolved because the record carries low confidence or a validation flag. Coded fee incidence is 22.6 per cent. Among the 493 resolved records it is 22.9 per cent. Treating every unresolved record as a charger raises it to 24.0 per cent.

\begin{figure}[htbp]
\centering
\includegraphics[width=\linewidth]{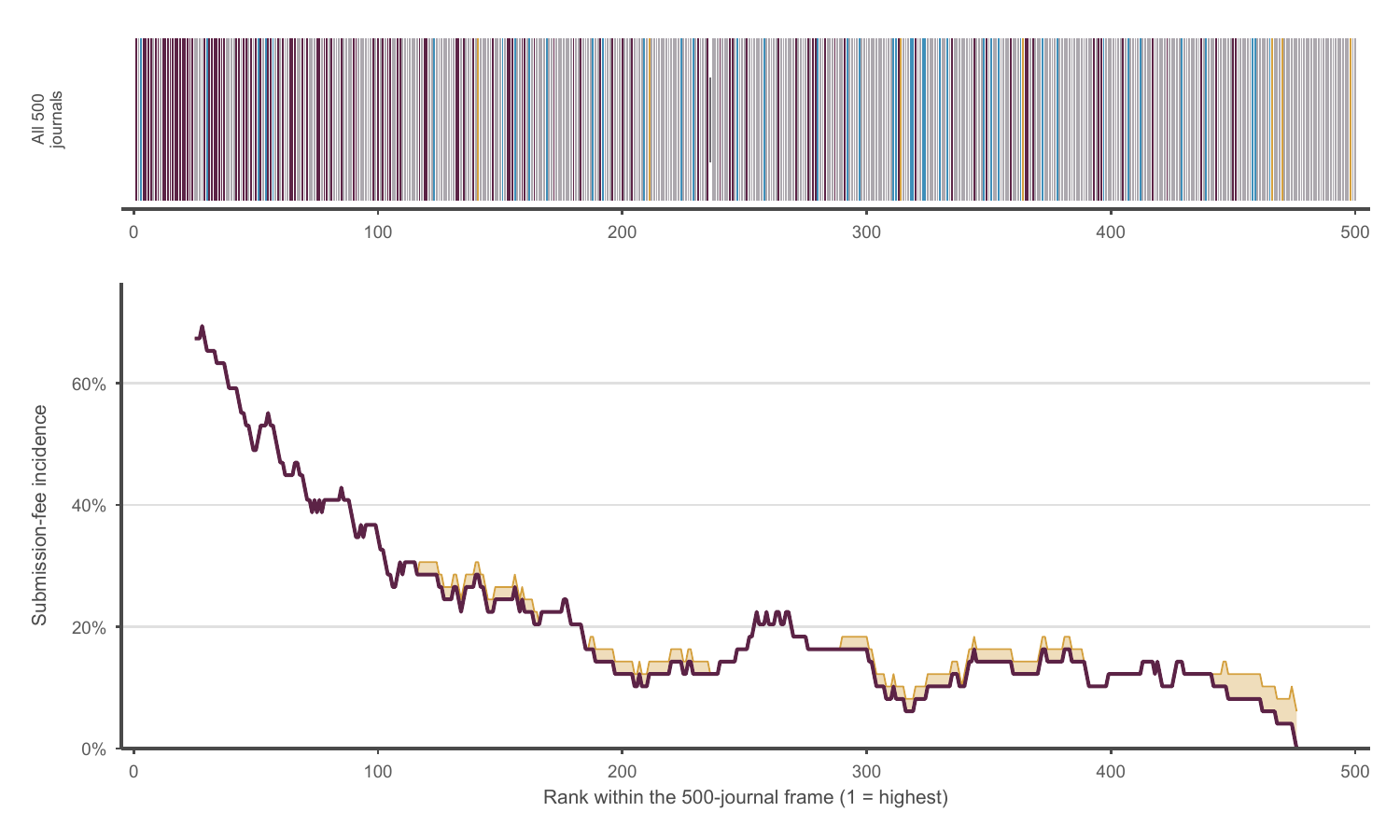}
\caption{\textbf{The census of submission fees, July 2026.} The upper panel draws one tick per journal in the frame, in rank order, coloured by evidence class: plum ticks charge for submission, blue ticks carry text stating that no submission charge applies, light grey ticks show no charge after the prescribed search, gold ticks are unresolved and the short dark tick admits only members. The lower panel plots the share charging (plum line) within a centred 49-journal window of the 500-journal frame. The gold band raises every unresolved record in the window to a charger. The unit of observation is a journal. The band measures classification sensitivity, not sampling uncertainty.}
\label{fig:incidence}
\end{figure}

Fee incidence falls steeply with rank (Figure~\ref{fig:incidence}). In the first fiftieth of the frame 34 of 50 journals charge a fee. One of 50 does so in the last. The coded shares are 68 and 2 per cent. The path between them is not monotonic. The overall gradient is nevertheless large. Among journals with an established price the median converted fee is \$143, over the 111 records whose amount and currency are both known. The mean among the top hundred is \$258, over the 54 chargers there. Ten records report referee payment in some form, 88 report a waiver for authors in low-income countries and 18 report a student rate. The referee-payment records cover heterogeneous arrangements. They do not show ten journals paying a market wage.

\subsection{Historical direction and publisher architecture}

A fee amount and a currency label are different objects. The 2019 comparison depends on that difference. It also depends on linking each journal in \citet{ko2019fees} to the right journal in the census, which is less mechanical than it sounds. Ranking entries carry the publisher's name after the title. A title match against the raw entry can therefore prefer a similarly named journal to the correct one. Sixty-six of the 69 journals in that source are linked here by a reviewed crosswalk, which records for each link whether it was exact. It also names the three journals deliberately left unlinked because the census does not contain them.

That source records the posted currency and whether a journal required membership. I therefore call a change in the level of a fee comparable only when both endpoints are positive amounts in the same posted currency. The \emph{European Economic Review}'s rise from 100 to 125 euros is comparable. A move between a euro amount and a dollar amount is not. One journal falls out of the comparison on exactly that ground. Introductions and removals stay comparable in direction because one endpoint is zero.

\begin{figure}[htbp]
\centering
\includegraphics[width=\linewidth]{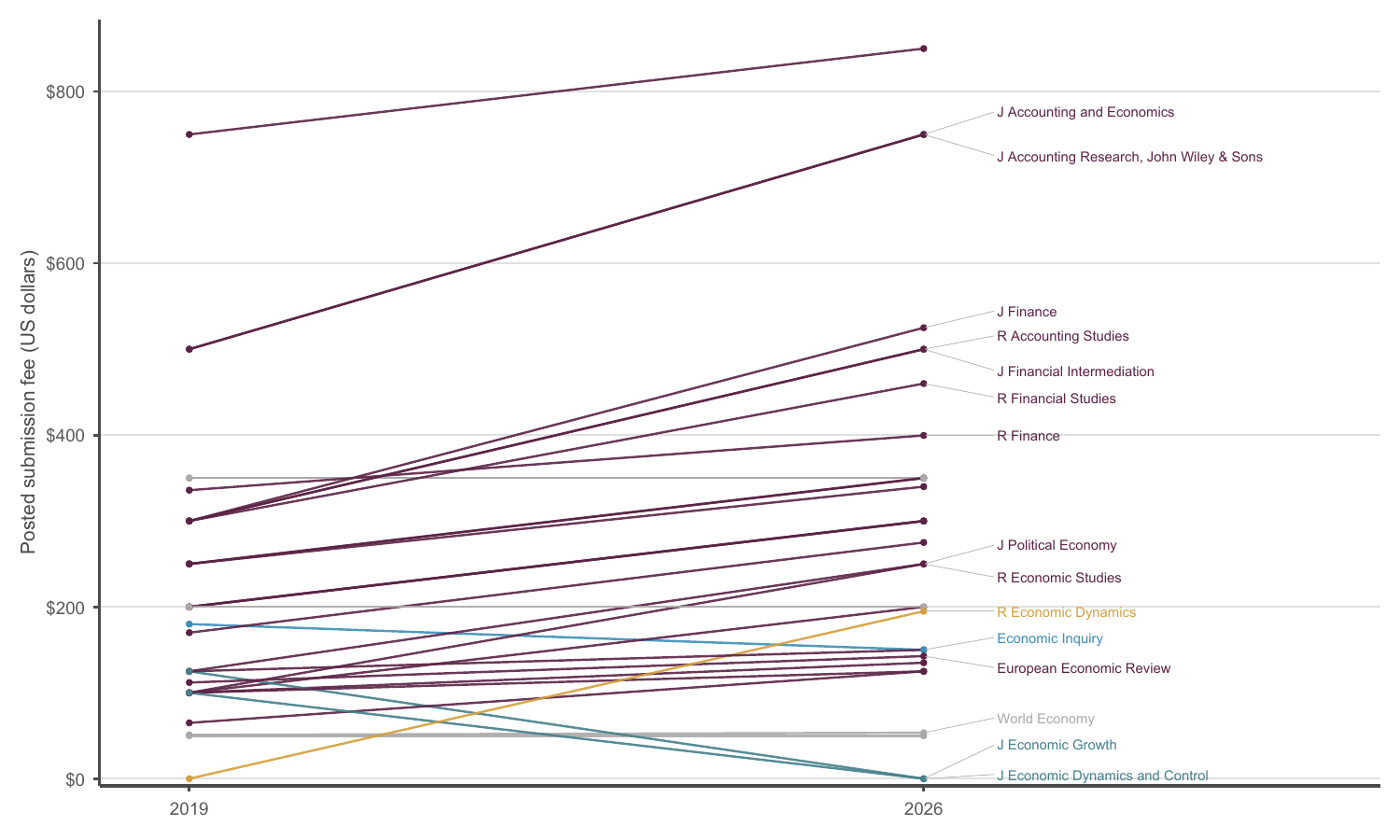}
\caption{\textbf{Posted submission fees in 2019 and 2026.} Each line is one of the 32 matched journals whose endpoints are directly comparable: both positive in the same posted currency, introduced from zero or removed to zero. Plum lines raised the fee, blue lines lowered it, grey lines left it unchanged, the gold line introduced a fee and teal lines removed one. Classes are assigned within the posted currency. For display, euro and sterling amounts are converted to dollars at European Central Bank reference rates: the 2019 annual averages of 1.12 and 1.28, and the census-month averages of 1.14 and 1.34 for July 2026. Twenty-eight further journals were at zero in both years and five required membership in 2019. One more changed its posted currency and cannot be compared. None of those groups is drawn. The unit of observation is a matched journal.}
\label{fig:history}
\end{figure}

Under that rule 23 journals raised a positive nominal fee in the same currency, one lowered it and five left it unchanged. One journal introduced an ordinary fee and two removed one. Twenty-eight remained at zero. A further five required membership in 2019. Membership buys more than the right to submit. It is therefore not a zero-dollar entry price. Those cases are reported separately. Figure~\ref{fig:history} traces each of the 32 comparable fees from its 2019 level to its 2026 level. The classification never compares across currencies. The display converts the euro and sterling fees to dollars at each year's exchange rate.

Pricing also differs across publisher portfolios. Fee incidence is 34 per cent at Elsevier and 32 per cent at Wiley, against 3 per cent at Springer Nature and 7 per cent at Taylor \& Francis. Table~\ref{tab:publisher} reports linear-probability models: regressions of an indicator for charging a fee on rank and journal characteristics. They are estimated on the 493 resolved records, because a record whose fee status could not be settled is not evidence of a zero and does not belong in the dependent variable. A one-unit rise in log rank is associated with a 13.3 percentage-point lower probability of charging after publisher fixed effects. The society-owned coefficient is indistinguishable from zero, with a 95 per cent interval from $-7.8$ to 5.2 points. Open-access status is unknown for a large minority of journals and enters as its own category rather than being treated as closed access. These are descriptions of a cross-section. Publisher composition, rank, ownership, submission demand and institutional subsidy are determined together. No coefficient has a causal reading.

\begin{table}[htbp]
\centering
\caption{\textbf{Submission-fee incidence, rank and publisher}}
\label{tab:publisher}
\begin{tabular}{lccc}
\toprule
 & (1) & (2) & (3) \\
\midrule
Log rank in frame & $-0.178$ & $-0.158$ & $-0.133$ \\
 & $(0.021)$ & $(0.022)$ & $(0.022)$ \\
Society owned &  & $-0.047$ & $-0.013$ \\
 &  & $(0.034)$ & $(0.033)$ \\
Open access &  & $-0.141$ & $-0.116$ \\
 &  & $(0.053)$ & $(0.057)$ \\
Open access unknown &  & $-0.159$ & $-0.174$ \\
 &  & $(0.035)$ & $(0.035)$ \\
Posts an APC &  & $-0.053$ & $-0.147$ \\
 &  & $(0.034)$ & $(0.039)$ \\
Publisher fixed effects & No & No & Yes \\
Observations & 493 & 493 & 493 \\
$R^2$ & 0.171 & 0.212 & 0.296 \\
\bottomrule
\end{tabular}
\begin{minipage}{0.94\textwidth}
\vspace{2pt}\emph{Notes:} Linear-probability models. The dependent variable equals one when the journal charges for submission. The unit of observation is a journal in the 500-journal frame. The sample is the 493 journals whose fee status is resolved, since an unresolved record is not evidence of a zero. Treating all 500 as resolved leaves the rank coefficient at $-0.133$ with the same standard error. Open-access status is unknown for a large minority of journals and enters as its own category rather than as closed access. Heteroskedasticity-consistent HC1 standard errors are in parentheses. Publisher fixed effects group portfolios with at least ten journals and combine smaller portfolios in an other category. Estimates are descriptive associations.
\end{minipage}
\end{table}

\section{Money, delay and submission cascades}\label{sec:model}

\subsection{The value of an attempt}

Journals form tiers $j=1,\ldots,J$ ordered by rank. Author $i$ at tier $j$ believes the manuscript will be accepted with probability $p_{ij}$. Acceptance is worth $V_j$: the value of publishing in that tier. If the author skips the tier, or continues after rejection, the manuscript goes to the next journal down. That continuation is worth $C_{i,j+1}$ from the moment it starts. A decision arrives only after an expected delay $d_j$. The author discounts the future at rate $\rho_i$ and bears a direct waiting cost $c_i$ per period. The journal charges an upfront fee $f_j$.

Measured before the fee and against proceeding immediately, the expected value of trying tier $j$ is
\begin{equation}
x_{ij}
=e^{-\rho_i d_j}\left[C_{i,j+1}
  +p_{ij}\left(V_j-C_{i,j+1}\right)\right]
  -C_{i,j+1}-c_i d_j.
\label{eq:attempt}
\end{equation}
The author submits when $x_{ij}\geq f_j$. Two features of equation~\eqref{eq:attempt} carry the argument. First, the prize from acceptance is not the whole value of publication. It is the difference $V_j-C_{i,j+1}$ between publishing here and publishing at the next journal down. A rejected paper is not lost. It moves on. Second, delay is already a price. Waiting shrinks both the prize and the fallback through discounting. It also consumes the author's time directly.

Let $G_j$ denote the distribution of $x_{ij}$ across the mass $M_j$ of manuscripts that could reach tier $j$ before any fee is charged. Raw submissions are
\begin{equation}
D_j(f_j)=M_j\left[1-G_j(f_j)\right].
\label{eq:demand}
\end{equation}
Where $G_j$ is atomless (no positive mass of authors shares exactly the same attempt value) the weak and the strict submission inequality give the same demand. Proposition~\ref{prop:residual} imposes that condition. The single number $x_{ij}$ absorbs many kinds of difference across authors: beliefs, career horizons, waiting costs, liquidity and outside options. A fee therefore selects the manuscripts whose authors privately value the attempt most. That is not the same as selecting the best science. The two coincide only if authors who value the attempt more also hold better manuscripts on average, formally if $\mathbb E[\theta_i\mid x_{ij}]$ increases in $x_{ij}$.

\subsection{Capacity in completed reports}

Every raw submission consumes some editorial attention. External referee capacity is a different and scarcer input. Let $q_j$ be the share of submissions sent beyond desk triage. Let $k_j$ be the mean number of completed reports per externally reviewed manuscript. Report demand is therefore $q_jk_jD_j$ rather than $D_j$. Let $A_j$ be the number of completed reports the journal can obtain from its referees.

\begin{proposition}[Residual monetary price]\label{prop:residual}
Suppose $G_j$ is continuous and atomless, $q_j,k_j>0$ and the desk-triage share is locally fixed. The non-negative fee that equates completed-report demand to capacity is
\begin{equation}
f_j^*=\begin{cases}
0, & q_jk_jM_j[1-G_j(0)]\leq A_j,\\[3pt]
G_j^{-1}\!\left(1-\dfrac{A_j}{q_jk_jM_j}\right),
& q_jk_jM_j[1-G_j(0)]>A_j,
\end{cases}
\label{eq:fee_general}
\end{equation}
and the second line is positive under the stated condition. If $G_j$ has an atom at the cutoff, exact clearing requires a lottery among the authors at that cutoff.
\end{proposition}

The proof substitutes equation~\eqref{eq:demand} into $q_jk_jD_j=A_j$. The atomless condition does real work. If a positive mass of authors sits exactly at the cutoff, demand jumps past the target as the fee crosses their value. A generalised inverse then identifies a cutoff and nothing more. Exact clearing needs a lottery that admits only part of the mass at the cutoff.

This fee is a queue-clearing benchmark, not a journal optimum. The model asks what money price would clear a given report constraint. A posted fee may differ because journals also value revenue, readership and mission and because report capacity may itself respond to payment.

The uniform-belief case shows in one line how delay changes the money price. Suppose $p_{ij}$ is uniform on the unit interval. Suppose $V_j$, $C_{i,j+1}$, $\rho_i$, $c_i$ and $d_j$ are common within the tier. Write $\Delta_j=V_j-C_{j+1}$ for the prize from placing here rather than at the next journal. Write $z_j=A_j/(q_jk_jM_j)$ for report capacity as a share of the report demand that would arise if every potential author submitted. Then
\begin{equation}
f_j^*=\left[
e^{-\rho d_j}\Delta_j(1-z_j)
-(1-e^{-\rho d_j})C_{j+1}-c d_j
\right]_+.
\label{eq:fee_uniform}
\end{equation}
In words: scarcity fixes the full price of entry. Delay collects part of that price. The fee is whatever remains. A journal whose authors already wait a year can clear the same queue with a smaller fee than a fast journal could. The result also matters for measurement. Inferring congestion from a posted fee alone treats the fee as the whole price. Delay collects part of the price. The inference therefore understates how congested the journal is.

The distinction between submissions and reports disciplines the funding arithmetic. Suppose every submitted manuscript pays $f_j$. Suppose all revenue is paid out to referees at an honorarium $w_j$ per completed report. Then
\begin{equation}
f_jD_j=w_jq_jk_jD_j,
\qquad w_j=\frac{f_j}{q_jk_j},
\label{eq:funding}
\end{equation}
and with a pooled reviewer budget the honorarium is total fee revenue divided by total completed reports. Equation~\eqref{eq:funding} is an accounting identity for a counterfactual. It is not an observed wage. Paying referees expands capacity only if referees are actually paid and only if their supply of reports responds to the payment.

\subsection{Routing and congestion export}

In a ranked system the potential mass $M_j$ is not fixed. Authors may skip an expensive tier. They may continue after rejection or give up. Rejection itself teaches authors something about their chances further down. Through both channels the demand facing each tier depends on the prices and decisions of the tiers above it.

A deliberately simple chain isolates the mechanism. Let every unsuccessful attempt lead to a further attempt with probability $r$. Allow at most $J$ tiers.

\begin{proposition}[Submission-chain multiplier]\label{prop:routing}
The expected number of submission attempts generated by one manuscript is
\begin{equation}
L_J(r)=\sum_{\ell=0}^{J-1}r^\ell.
\label{eq:multiplier}
\end{equation}
It rises from one when every manuscript exits after its first attempt to $J$ under complete continuation. A local fee reduces report demand for the system only through the authors who exit, shorten their chain or move to tiers with lower report intensity. If every excluded author reroutes immediately to an otherwise identical next tier, the fee transfers the next attempt instead of eliminating it.
\end{proposition}

The result follows by summing the probability of reaching each tier. Its point is the difference between clearing one queue and relieving the system. A fee that only diverts a manuscript to the next journal does not reduce the total number of reports the profession must write. The same distinction applies to speed. Faster decisions cut the author's waiting cost. They also deliver rejected manuscripts to the next journal sooner. A full quantitative answer needs two objects this paper does not observe: the matrix of moves between tiers and the way beliefs respond to rejection. Holding both fixed turns any projection into a partial-equilibrium pressure calculation.

\section{Quantitative scenarios}\label{sec:quantitative}

What current data can support is limited. The limits shape this section. The census identifies posted fees and institutional menus. It does not identify the distribution of attempt values $G_j$, potential arrivals $M_j$, report capacity $A_j$, the routing of rejected papers or the response of referee supply to payment. Published editor reports fill in fragments. Among the top 200 entries, 53 have usable reports, 35 state annual submissions, 20 state a desk-rejection rate, 34 state a turnaround measure and only two state a total number of referee reports. Two report counts are not enough to convert submissions into reports across the ranking.

Two further gaps close off a dollar answer. A fee schedule inferred from posted fees and then compared with the same posted fees fits by construction. It cannot test itself. The salary gain associated with a top publication is the second gap. It measures the gross difference in pay between people who publish there and people who do not. Equation~\eqref{eq:attempt} needs something narrower: the value of this acceptance over the author's next-best placement. I therefore report the model's quantities as shares of that placement prize rather than in dollars.

\begin{figure}[htbp]
\centering
\includegraphics[width=\linewidth]{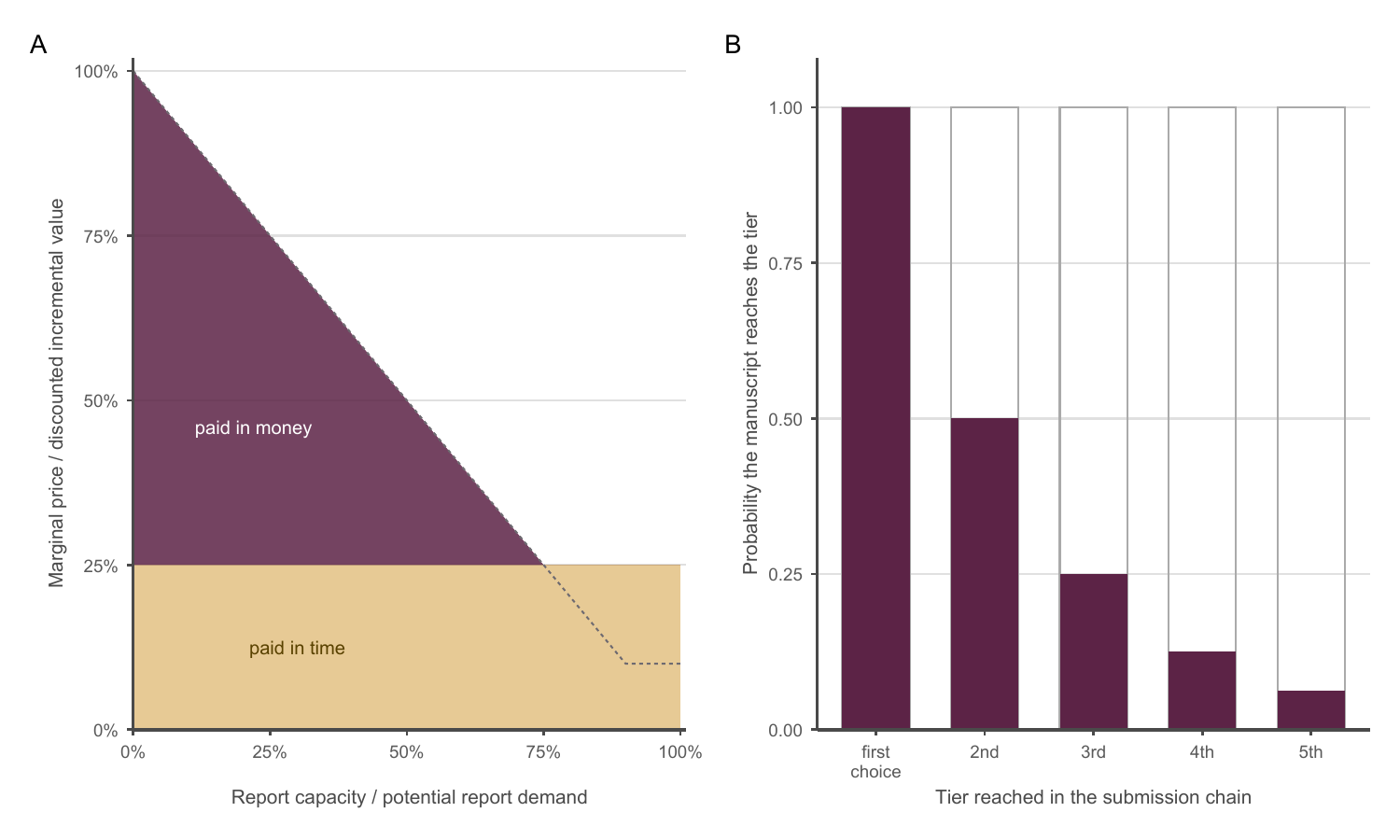}
\caption{\textbf{Money, delay and routing in the benchmark model.} Panel A shows the price that clears report capacity under uniform beliefs, as a share of the discounted incremental value of placement. The marginal entrant always bears the delay burden (gold), set at 25 per cent of that value. The money fee (plum) is the residual above the delay floor and reaches zero while capacity is still scarce. The dashed line repeats the schedule for a 10 per cent burden. The horizontal axis is report capacity divided by the report demand that every potential manuscript would generate. Panel B shows the probability that one manuscript reaches each of five tiers when half of rejected authors continue after each rejection (solid bars), against complete continuation (outlined bars). Expected attempts per manuscript, equation~\eqref{eq:multiplier}, are 1.94 and five. Both panels are theoretical scenario maps. No line is fitted to posted fees.}
\label{fig:modelmap}
\end{figure}

Figure~\ref{fig:modelmap} shows what the model delivers without a dollar scale. Panel A takes the uniform-belief case. Write $\tilde\Delta_j=e^{-\rho d_j}\Delta_j$ for the prize discounted back over the waiting period. Let
\[
h_j=\frac{(1-e^{-\rho d_j})C_{j+1}+c d_j}{\tilde\Delta_j}
\]
be the share of that discounted prize that waiting already costs the marginal author. Equation~\eqref{eq:fee_uniform} then becomes
\[
\frac{f_j^*}{\tilde\Delta_j}=\left[1-z_j-h_j\right]_+ .
\]
The reading is direct. As report capacity rises toward potential demand the full price of entry falls. The delay burden does not move. The fee is the layer between the two. It reaches zero while capacity is still scarce, at the point where waiting alone deters enough authors. Panel B plots the chain of expected attempts for at most five tiers. A continuation probability of one half produces 1.94 expected attempts per manuscript. Complete continuation produces five. Both panels map assumptions into outcomes. Neither estimates a parameter.

\begin{figure}[htbp]
\centering
\includegraphics[width=\linewidth]{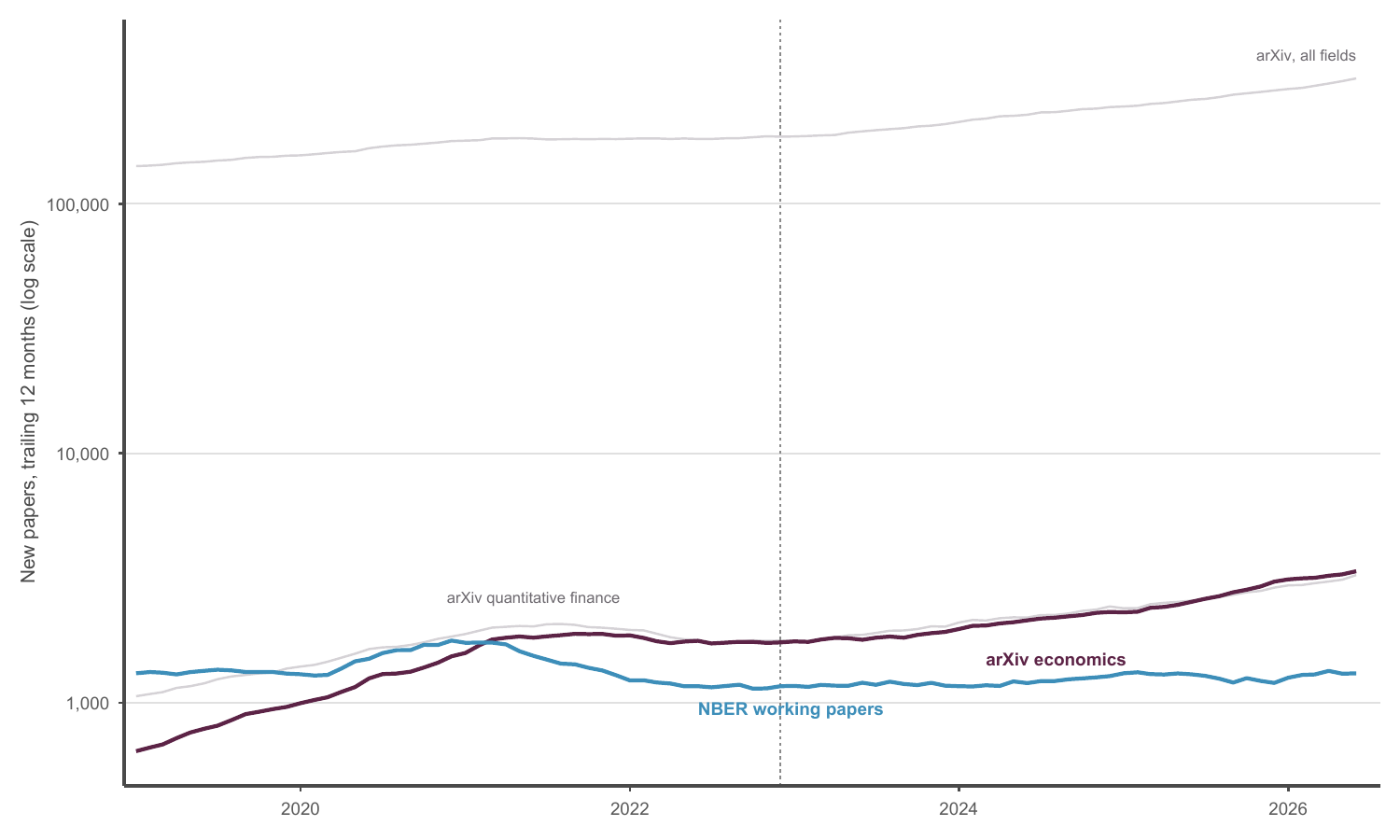}
\caption{\textbf{Manuscript growth where posting is open and where it is gated.} Twelve-month trailing totals of new papers, log scale. The dashed line marks the release of ChatGPT. In the year to June 2026 arXiv economics grew 0.28 log points while the affiliation-gated NBER series was approximately flat. The grey series show arXiv in all fields and in quantitative finance for scale. The unit is a posted paper rather than a journal submission or a report request.}
\label{fig:growth}
\end{figure}

Observed manuscript series still show the direction of pressure. In the year to June 2026 trailing arXiv economics postings grew by 0.28 log points while the affiliation-gated NBER series was approximately flat (Figure~\ref{fig:growth}). Growth appears where posting is open and not where posting is gated. Postings are neither journal submissions nor submission attempts. The contrast therefore does not identify $M_j$ or its future path. \citet{kwa2025metr} report that the length of software tasks AI systems complete at fifty per cent reliability doubles about every seven months. A capability clock is not a forecast of journal arrivals. I do not project fees from it.

An independently anchored dollar exercise needs five things the profession could collect. Journal administrative data would measure raw submissions, desk survival, reports per reviewed manuscript and decision delay. Observed fee changes, or menus that let authors trade money against waiting time, would identify the distribution of attempt values. Author submission histories would identify where rejected papers go and how beliefs respond to rejection. Variation in referee payment would identify whether paid referees complete more reports. And placement comparisons would pin down the prize $V_j-C_{j+1}$ that a gross salary association cannot. Until then, equation~\eqref{eq:fee_general} is best used to report bounds and break-even combinations.

\section{Institutional design}\label{sec:design}

\subsection{Waivers and liquidity}

A waiver changes demand and revenue at the same time. Let a share $s$ of potential authors be eligible for the discounted price $\alpha f$ with $\alpha<1$. Let $G_E$ and $G_N$ be the attempt-value distributions of eligible and non-eligible authors. Raw demand becomes
\begin{equation}
D(f,\alpha)=M\left[(1-s)\{1-G_N(f)\}
 +s\{1-G_E(\alpha f)\}\right],
\label{eq:waiver_demand}
\end{equation}
and fee revenue is $fM(1-s)\{1-G_N(f)\}+\alpha fMs\{1-G_E(\alpha f)\}$. The practical implication is that a waiver cannot be costed by simple subtraction. Cutting the fee for one group changes how many authors submit, how many reports are needed, what fee clears the market and what revenue remains. These objects move together and must be solved together.

Whether a waiver improves the quality of the submission pool depends on the marginal entrant, not on the average member of the eligible group. A discount admits the authors who were just priced out. Mean quality rises only if those marginal authors carry better manuscripts than the existing pool does. The 88 coded low-income-country waivers show that tiered menus are administratively feasible. The census contains no author-level beliefs, income or manuscript quality. It cannot say whether the marginal entrants are strong. The equity concern documented by \citet{oldford2023marginalizing} therefore remains open as a question about who bears the price.

\subsection{Payment timing}

When the fee is paid matters as much as how large it is. An upfront non-refundable fee produces the cutoff in equation~\eqref{eq:attempt}: authors submit when the attempt is worth at least the fee. A deferred fee owed regardless of outcome preserves that cutoff when its present value is held fixed and payment can be enforced. It also spares authors who lack cash today. A rebate on desk rejection looks similar but is not. The expected fee then depends on each author's own guess about desk rejection. The instrument therefore charges a different price to every author. A fee collected only on acceptance changes the problem more deeply. It replaces the acceptance term $e^{-\rho d}p\Delta$ with $e^{-\rho d}p(\Delta-f)$. With no time cost, every author with a positive belief submits whenever the incremental prize exceeds the charge. In the full model waiting can still screen authors, but the acceptance-contingent fee adds no fixed hurdle to an attempt. Collecting the full fee upfront and refunding it later helps no author who lacks the cash today. These four instruments are different contracts and should be evaluated as such.

\subsection{Triage and verification technology}

AI-assisted triage enters the model through $q_j$. Better early screening lowers the share of submissions that reach referees. Verification tools can instead raise effective capacity $A_j$. The current evidence supports a bounded role for both. \citet{pataranutaporn2025peerreview} find that large language models separate stronger from weaker economics papers. The same evaluations show author-name and affiliation effects when the reviewing model sees identities. In finance, \citet{newton2026aireferees} also find useful separation. Their targeted probes report scores largely insensitive to prompt injection and author affiliation. They find conservative scoring at the top of the quality range and retain a role for expert judgement. The two studies agree on separation. They differ on sensitivity to author identity.

The institutional choice therefore runs across five margins: a money price, a time price, desk triage, paid reports and investment in verification. A fee controls how much arrives. It does not ensure that what arrives is reviewed carefully. Triage lowers $q_j$. Added referee capacity raises $A_j$. Both reduce the fee needed in equation~\eqref{eq:fee_general}. What each does to the whole cascade depends on how quickly rejected authors continue and on what rejection teaches them.

\section{Conclusion}\label{sec:conclusion}

What is the price of submission? For an author it has three parts: the fee paid on submission, the months spent waiting for a decision and the option given up while waiting, namely trying the next journal immediately. These parts substitute for one another. A journal that makes authors wait needs less money to ration the same queue. A rejected manuscript usually moves on to another journal. A fee at one journal therefore shortens its own queue without necessarily shortening any other. The system saves referee work only when a fee persuades some authors to stop trying, to try fewer journals or to move to journals whose review uses fewer reports. When a fee does none of these things it passes congestion down the ranking.

The census makes the money part of that price visible. Of the 500 highest-ranked economics journals that accept unsolicited submissions, 113 charge for submission, a share of 22.6 per cent. Treating every unresolved record as charging raises the classification bound to 24.0 per cent. The slope matters more than the level: 68 per cent of the highest-ranked fiftieth charges a fee against 2 per cent of the lowest. In the strict nominal comparison with 2019, 23 positive same-currency fees rose, one fell, two were removed and one was introduced. Explicit prices are concentrated at the top of the ranking. Among the positive fees that can be compared over time, most increased.

The model warns against two natural readings of these facts. The first reads a journal's fee as a measure of its congestion. That reading fails because waiting time is also a price. A journal that takes a year to decide imposes a large cost on its authors even when its fee is zero. A low fee can therefore coexist with severe rationing in time. The second reading divides fee revenue by submissions and calls the result a potential referee payment. That reading fails on units. Only some submissions reach referees. Each of those needs several reports. The payment per completed report is the fee divided by completed reports per submission, so it may lie below or above the fee. What the data do support is a conditional statement: given report capacity, the delay burden and the destination of rejected papers, the model says how much of the entry price must be collected in money.

These results connect the price schedule to the wider evidence on scholarly publishing. \citet{hanson2024strain} document aggregate pressure on the publishing system; the economics census shows that journals distribute some of that pressure through explicit entry prices concentrated at the top, while delay bears the rest. Evidence that author-facing charges stratify access \citep{klebel2023barrier,oldford2023marginalizing} makes waiver design part of capacity policy rather than an afterthought, although the census cannot identify which authors are deterred. On the supply side, evidence that referee effort varies \citep{zhang2022endeavor} and that additional review is not consistently associated with more citations \citep{wolfram2021exploration} explains why report counts and report value must remain separate. The model therefore counts completed reports without claiming that any particular number is optimal. It describes a system in which authors submit to one journal at a time and report capacity is fixed. Simultaneous submission would change the queue-shifting result. Referee payment would change it too if compensation expands report capacity.

The open question is what happens to the manuscripts a fee turns away. Posted prices cannot answer it. Journal records around actual fee changes could: how many manuscripts arrive, how many pass the desk, how many reports they consume, how long decisions take and where rejected papers appear next. Data of that kind would show whether a submission fee removes weak attempts from the system or only reroutes them. That difference decides whether a fee buys the profession more careful review or buys one journal a shorter queue.

\bibliographystyle{aea}
\bibliography{references}

\appendix
\renewcommand{\thesection}{A\arabic{section}}
\setcounter{section}{0}

\section{Data and verification}\label{sec:appendix}

\textbf{The frame.} The ranking snapshot is pinned to July 2026 and the raw collection is untouched. Candidates for removal are nominated from the RePEc archive code, from anchored title patterns, from the absence of a registered identity in Crossref or Scimago and from the absence of any deposited article since 2023. Each candidate is then resolved on its own pages. The retained verdict, quotation and address are recorded in \path{data/frame_classification.csv}. Nomination alone never removes an entry. Four entries were settled by reading the RePEc series page directly, because a search can answer about a different publication with the same generic title. Journals that publish only commissioned work carry their own flag rather than a silent removal. The frame can be recomputed with or without them.

\textbf{Current-fee evidence.} The evidence script classifies each record from retained fields and leaves the raw files untouched. Charging, price and evidence quality are separate variables. A record can therefore charge with an unrecovered amount without being counted as a zero. A record whose status is unresolved is never counted as a zero either. The explicit-zero rule requires retrieved text in which a negation and a submission-specific phrase occur in the same clause. Text about an article-processing charge, a page charge or membership dues is not evidence about a submission fee in either direction. Membership decides the class only when no per-submission amount stands behind it, which is the case when no amount was recovered or when members submit free. Rows are in \path{data/revision_fee_evidence_audit.csv}.

\textbf{Historical comparison.} Journals in \citet{ko2019fees} are linked to the census by a reviewed crosswalk, \path{data/ko_crosswalk.csv}, which matches each title against both the raw ranking entry and the entry with its trailing publisher clause removed, assigns links one to one and records the method for every link. Sixty-two links are exact, one is accepted on similarity and three were decided by hand. Three journals are left unlinked because the census does not contain them. Positive amounts are compared only within the same posted currency. A move from an ordinary zero to a positive fee is an introduction and the reverse is a removal. A mandatory-membership requirement is neither. Conversions for display use European Central Bank reference rates from \path{data/raw/fx_rates.csv}. The 2019 annual averages are 1.12 dollars per euro and 1.28 per pound. The July 2026 averages are 1.14 and 1.34.

\textbf{Regression.} Table~\ref{tab:publisher} is reproduced by \texttt{code/08\_revision\_evidence.R}. The script canonicalises publisher names, estimates on resolved records within the frame, treats unknown open-access status as its own category and computes HC1 standard errors from the linear-model design matrix. A specification that instead treats every unresolved record as a zero is reported alongside as a bound. No coefficient is interpreted causally.

\textbf{Model verification.} Twenty-two numerical checks reproduce the attempt-value equation, the clearing condition in report units, the revenue-equivalent honorarium, both slack corners, the direction of the bias from ignoring delay, the normalised residual-fee identity and the delay-money decomposition behind Figure~\ref{fig:modelmap}, cutoff randomisation with an atom in beliefs, the five-tier routing multiplier and the difference between upfront and acceptance-contingent payment. The checks are in \path{code/08_verify_revision_model.R}. All four figures are produced by \path{code/08_revision_figures.R}.

\end{document}